\documentclass[aps,prb,twocolumn,superscriptaddress]{revtex4}
\usepackage{graphicx}
\usepackage{amsmath}
\usepackage{amssymb}
\usepackage{amsfonts}
\usepackage{time}

\newcommand{\vect}[1]{{\mathbf #1}}
\newcommand{\beq}{\begin{equation}}
\newcommand{\eneq}{\end{equation}}
\newcommand{\braket}[1]{{\langle #1 \rangle}}

\begin{document}

\title{Experimental Consequences of the S-wave $\cos(k_x) \cdot \cos(k_y)$
Superconductivity in the Iron-Pnictides}

\author{Meera M. Parish}
\affiliation{Department of Physics, Princeton University, Princeton,
NJ 08544} %
\affiliation{Princeton Center for Theoretical Science, Princeton
University, Princeton, NJ 08544}

\author{Jiangping Hu}
\affiliation{Department of Physics, Purdue University, West
Lafayette, IN 47907}

\author{B. Andrei Bernevig}
\affiliation{Department of Physics, Princeton University, Princeton,
NJ 08544} %
\affiliation{Princeton Center for Theoretical Science, Princeton
University, Princeton, NJ 08544}

\date{\now,
\today}

\begin{abstract}
The experimental consequences of different order parameters in
iron-based superconductors are  theoretically analyzed. We consider
both nodeless and nodal order parameters, with an emphasis on the
$\cos(k_x)\cdot \cos(k_y)$ nodeless order parameter recently derived
by two of us~\cite{seo2008}. We analyze the effect of this order
parameter on the spectral function, density of states, tunneling
differential conductance, penetration depth, and the NMR spin
relaxation time. This extended $s$-wave symmetry has line-zeroes in
between the electron and hole pockets, but they do not intersect the
two Fermi surfaces for moderate doping, and the superconductor is
fully gapped. However, this suggests several quantitative tests: the
exponential decay of the penetration depth weakens and the density
of states reveals a smaller gap upon electron or hole doping.
Moreover, the $\cos(k_x) \cdot \cos(k_y)$ superconducting gap is
largest on the smallest (hole) Fermi surface. For the $1/T_1$ NMR
spin relaxation rate, the inter-band contribution is consistent with
the current experimental results, including a (non-universal)
$T^{3}$ behavior and the absence of a coherence peak. However,  the
intra-band contribution is considerably larger than the inter-band
contributions and still exhibits a small enhancement in the NMR spin
relaxation rate right below $T_c$ in the clean limit.
\end{abstract}



\maketitle

\paragraph*{Introduction --}
The recent discovery of iron-based superconductors with a transition
temperature as high as 55K has stimulated a flurry of experimental
and theoretical
activity~\cite{kamihara2008,takahashi2008,ren2008,GFchen2008,XHchen2008,wen2008,rotter2008,sasmal2008,GFchen2008_3,Cvetkovic2008}.
However, a conclusive observation of the pairing symmetry still
remains elusive, with both nodal and nodeless order parameters
reported in experimental observations.

Numerical and analytic research suggests that the antiferromagnetic
exchange coupling between Fe sites is strong~\cite{Ma,yildirim,si}.
Owing to As-mediated hopping, antiferromagnetic exchange exists not
only between the nearest neighbor (NN) Fe sites, but also between
next nearest neighbor (NNN) sites. Moreover, the NNN coupling
strength $J_2$ is stronger than the NN coupling strength $J_1$. The
$J_1-J_2$ model produces half-filled magnetic physics consistent
with experimental neutron data~\cite{mook}. A nematic magnetic phase
transition has been predicted in this model~\cite{Fang2008,Xu2008},
consistent with the experimental observation of a structural
transition preceding the spin density wave (SDW) formation. This
model suffers, however, from an important deficiency -  it is an
insulator whereas the real material is an, albeit bad, metal. We,
however, believe that the spin-spin interaction insight is important
to the physics of the iron-pnictides.

In a recent paper \cite{seo2008}, two of us added electron itineracy
to the problem and studied  a $t-J_1-J_2$ model \emph{without} band
renormalization. We found that the singlet-forming $J_1-J_2$
interaction gives rise to four possible pairing symmetries:
$\cos(k_x) \pm \cos(k_y)$, $\sin(k_x)\cdot \sin(k_y)$ and $\cos(k_x)
\cdot \cos(k_y)$. The last two are strongly preferred from an
interaction standpoint when $J_2>J_1$, but only $\cos(k_x) \cdot
\cos(k_y)$ matches the symmetry of the iron-pnictide Fermi surface:
it is maximal around $(0,0), (\pi,0), (0,\pi), (\pi,\pi)$ - the
location of the Fermi surfaces in the unfolded one-iron per site
Brilloiun zone. Although we used a specific, two band model for our
calculation \cite{seo2008}, our results are completely independent
of any model, \emph{as long as} the dominating interaction is
next-nearest neighbor $J_2$ \emph{and} the Fermi surfaces are
located close to the aforementioned spots in the Brillouin zone.
Some order parameters (such as $d_{xy} =\sin(k_x) \cdot \sin(k_y)$
and others) mismatch the Fermi surface symmetry and can be
discarded. We note that $\cos(k_x) \cdot \cos(k_y)$ changes sign
between the electron and hole pockets in the Brillouin zone. In this
sense, it resembles the order parameter proposed by Mazin through
weak-coupling general arguments \cite{mazin2008}. At moderate
doping, our gap is isotropic within the same Fermi surface, while
changing sign between electrons and hole pockets, but at relatively
high doping $\cos(k_x) \cdot \cos(k_y)$ exhibits some anisotropy
even within the same Fermi surface.

Neutron measurements have found antiferromagnetic stripe order of Fe
moments ranging from $0.26 \mu_B$ in NaOFeAs\cite{Dai2008b} and
$0.36\mu_B$\cite{Dai2008} in LaOFeAs to $0.8\mu_B$ in
CeOFeAs\cite{Dai2008a} and SrFe$_2$As$_2$~\cite{Dai2008c}.  A
magnetic moment of $0.8\mu_B$ is fully consistent with a purely
localized spin-one Heisenberg model. While a magnetic moment of
$0.3\mu_B$ is smaller than what is expected in a purely localized
spin-one system, it is rather  larger than what can be obtained in a
truly weak coupling theory. We point out that, due to imperfect
nesting, weak coupling theory requires large values of $U/t  \sim 4
$ to explain even small magnetic moments ($<0.2 \mu_B$), clearly
outside the weak-coupling limit~\cite{raghu2008}. Considering these
facts, together with the rather high resistivity of the
iron-pnictides, we find that the experimental evidence paints a
picture of the iron-pnictides as being at moderate interaction
couplings. Thus, moderate to strong coupling models can provide an
accurate qualitative description of the observed phenomena. In fact,
the $t-J_1-J_2$ model predicts the right physics of the parent state
SDW as well as the $\cos(k_x) \cdot \cos(k_y)$ order parameter.

\begin{figure}
\begin{center}
\includegraphics[width = 0.4\textwidth]{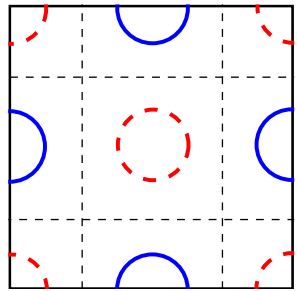}
\caption{\label{fig:fermisurf} Schematic diagram of the Fermi
surfaces in the iron-pnictides at half-filling in the unfolded
Brillouin zone $-\pi \leq k_x \leq \pi$, $-\pi \leq k_y \leq \pi$.
The dashed (red) and solid (blue) curves correspond to the hole and
electron Fermi surfaces, respectively. The dashed lines mark the
nodal lines at $(\pm \pi/2, k_y)$ and $(k_x, \pm \pi/2)$ for the
$\cos(k_x)\cdot\cos(k_y)$ order parameter proposed in
Ref.~\onlinecite{seo2008}.}
\end{center}
\end{figure}

In this paper we focus on the experimental properties of several
superconducting order parameters proposed in  the iron-pnictides,
with particular emphasis on the $\cos(k_x)\cdot \cos(k_y)$ order
parameter. We look at a simplified two-band superconducting model
and obtain the spectral function, density of states, tunneling
differential conductance, penetration depth and NMR spin relaxation
time. We stress the important point that the $\cos(k_x)\cdot
\cos(k_y)$ order parameter features lines of zeroes at $(\pm \pi/2,
k_y)$ and $(k_x,\pm \pi/2)$, as in Fig.~\ref{fig:fermisurf}
(obviously, irrespective of its harmonic form, any order parameter
changing sign between the electron and hole Fermi surfaces must have
 zero-lines). Thus, at low doping, the hole and electron Fermi
pockets are far away from the zero lines of the order parameter and
the superconductivity is nodeless.

Close to half-filling, we find that the $\cos(k_x)\cdot \cos(k_y)$
order parameter exhibits an exponentially decaying $\delta
\lambda(T) = \lambda(T)-\lambda(0)$, where $\lambda(T)$ is the
penetration depth at temperature $T$, as expected for a nodeless
superconductor. However, upon doping, the gap on the Fermi surface
varies in magnitude: for electron doping, the gap decreases on the
electron pocket and increases on the hole pocket. The penetration
depth is sensitive to the smallest gap in the system, and hence
exhibits a weakened exponential decay upon doping. This could
explain the conflicting values of the gap parameters obtained by
fitting the penetration depth experiments to the BCS exponential
form~\cite{malone2008,hashimoto2008,martin2008}. In the unlikely
event that the system remains superconducting at very large doping,
then the Fermi surfaces will cross the line of zeros of $\cos(k_x)
\cdot \cos(k_y)$ at around $35\%$ doping, and cause $\delta
\lambda(T)$ to become linearly dependent on T.

We also calculate the NMR spin relaxation rate $1/T_1$ of the bare
superconductor and find that it factorizes into inter- and
intra-band contributions. While, for the $\cos(k_x)\cdot \cos(k_y)$
order parameter, the inter-band contribution to the NMR spin
relaxation rate does \emph{not} exhibit a coherence peak, the
intra-band contribution is larger than the inter-band contribution
and still exhibits an enhancement right below $T_c$ owing to its
fully gapped $s$-wave nature. Adding the two contributions we find
that, although the coherence peak for $\cos(k_x)\cdot \cos(k_y)$ is
smaller than for a sign-preserving gap such as, for example,
$|\cos(k_x) \cdot \cos(k_y)|$, it is still present due to the
intra-band contribution. The coherence peak can be strongly reduced
if the intra-band scattering is stronger than inter-band scattering
or if the samples are strongly disordered. If the As structure
factor $A(\vect{q})$ is taken into account, the inter-band
contribution is severely reduced due to the fact that $A(\vect{q}) =
\cos(q_x/2 ) \cos(q_y/2)$ is zero close to the wavevector difference
between the electron and hole Fermi surfaces: $\vect{q} =(\pm
\pi,0), (0,\pm \pi)$. The As structure factor also reduces the
overall coherence peak by smearing the intra-band contribution.

\begin{figure}
\begin{center}
\includegraphics[width = 0.5\textwidth]{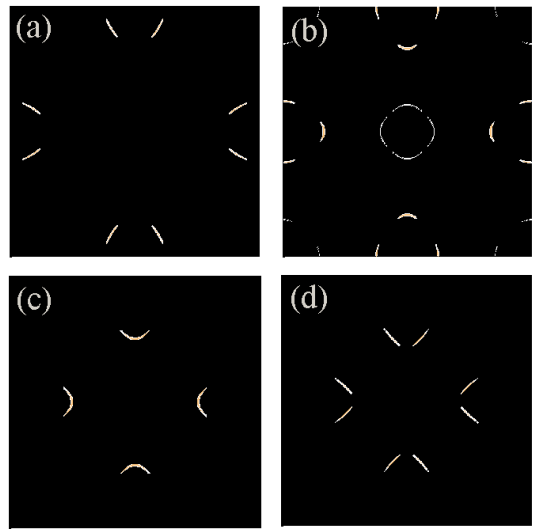}
\caption{\label{fig:Sfuncs} Behavior of the spectral function
$A(\vect{k},w)$ in an interval about the Fermi energy ($-0.02 < w <
0.02$) across the unfolded Brillouin zone $-\pi \leq k_x \leq \pi$,
$-\pi \leq k_y \leq \pi$ for gap parameter $\Delta_0 = 0.1$. Panels
(a) and (b) depict the $s_{x^2 + y^2}$ and $d_{xy}$ order
parameters, respectively, both at chemical potential $\mu = 1.6$.
The $s_{x^2y^2}$ order parameter is shown in panels (c) and (d) for
the higher electron doping values $\mu = 2$ and $\mu = 2.2$,
respectively. For these high doping values, the $s_{x^2 y^2}$
superconductor has become nodal. The lighter regions illustrate the
ungapped portions of the Fermi surface.}
\end{center}
\end{figure}

\paragraph*{Model --}
We approximate the typical iron-based material by a two-dimensional
square lattice of Fe atoms, since the superconductivity has been
shown to be associated with the FeAs layer. To capture the
degeneracy of the $d_{xz}$ and $d_{yz}$ orbitals on the Fe atoms, we
use the two-orbital per site model proposed by
Ref.~\onlinecite{raghu2008}. Although this description is only valid
in the case of an unphysically large crystal field splitting, we
particularize to this model for analytic simplicity. The kinetic
part of the Hamiltonian is written:
\begin{align}\label{eq:KE}
H_0 & =  \sum_{\vect{k}\sigma} \psi^\dag_{\vect{k}\sigma} \left(
\begin{array}{cc}
\epsilon_x(\vect{k}) - \mu & \epsilon_{xy}(\vect{k}) \\
\epsilon_{xy}(\vect{k}) & \epsilon_y(\vect{k}) - \mu
\end{array}
 \right) \psi_{\vect{k}\sigma}
\end{align}
Here, $\psi^\dag_{\vect{k}\sigma} = (c_{1,\vect{k},\sigma}^\dagger,
c_{2,\vect{k},\sigma}^\dagger)$ is the creation operator for spin
$\sigma$ electrons in the two orbitals $(1,2)= (d_{xz},d_{yz})$, $\mu$ is the chemical
potential, and the matrix elements are
\begin{eqnarray}\notag
  & \epsilon_x(\vect{k}) = -2t_1 \cos k_x - 2t_2 \cos k_y - 4t_3
  \cos k_x \cos k_y & \\ \notag
&\epsilon_y(\vect{k}) = -2t_2 \cos k_x - 2t_1 \cos k_y - 4t_3
  \cos k_x \cos k_y & \\
  & \epsilon_{xy}(\vect{k}) = -4t_4 \sin k_x \sin k_y &
\end{eqnarray}
While Eq.~\eqref{eq:KE} is only a simplified version of the true
band structure of the material, it produces Fermi pockets that
resemble those predicted by density functional theory (see
Fig.~\ref{fig:fermisurf}). The eigenvalues of \eqref{eq:KE} are:
\begin{equation}
E_\pm = \epsilon_+ - \mu \pm \sqrt{\epsilon_-^2 + \epsilon_{xy}^2}
\end{equation}
where $\epsilon_\pm = (\epsilon_x \pm \epsilon_y)/2$. In the
following, we take $t_1 = -1$, $t_2 = 1.3$, and $t_3 = t_4 = -0.85$.
The undoped compound, where there are two electrons per site,
corresponds to $\mu = 1.54$.

We now assume that the interacting part of the Hamiltonian induces
singlet pairing between electrons within each orbital, but we make
no further assumptions about the form of the interaction or the
pairing mechanism. Then we introduce pairing gaps $\Delta_{1,2}$ for
each orbital and we write down the mean-field effective Hamiltonian
$H(\Delta_1, \Delta_2) = \sum_\vect{k} \Psi(\vect{k})^\dagger
B(\vect{k})\Psi(\vect{k})$, where
\begin{equation} \label{eq:Bmat}
   B(\vect{k})=  \left(\begin{array}{cccc}\xi_x(\vect{k}) & \Delta_1(\vect{k}) & \epsilon_{xy}(\vect{k}) & 0 \\
                               \Delta_1^\ast(\vect{k}) & -\xi_x(\vect{k}) & 0 & -\epsilon_{xy}(\vect{k}) \\
                               \epsilon_{xy}(\vect{k}) & 0 & \xi_{y}(\vect{k}) & \Delta_2(\vect{k}) \\
                               0 & -\epsilon_{xy}(\vect{k}) & \Delta_2^\ast(\vect{k}) & -\xi_y(\vect{k})
             \end{array}\right)
\end{equation}
with $\xi_{x} = \epsilon_{x} - \mu$, $\xi_{y} = \epsilon_{y} - \mu$,
and we have used the four-component spinor $\Psi(\vect{k}) = (
c_{1,\vect{k},\uparrow}, c_{1,-\vect{k},\downarrow}^\dagger,
c_{2,\vect{k}, \uparrow}, c_{2,-\vect{k},\downarrow}^\dagger )$. We
neglect inter-orbital pairing in order to make the problem
analytically tractable. This is also reasonable because two of
us proved in Ref.~\onlinecite{seo2008} that, at least for the case
of the $t-J_1-J_2$ model (and hence for the most important gap we
will be focusing on - $\cos(k_x)\cdot \cos(k_y)$), the inter-orbital
pairing expectation value is negligible even in the case of strong
Hund's rule coupling.

The symmetry of the superconducting order parameter
$\Delta(\vect{k})$ has two possible $d$-wave types \cite{seo2008}:
$d_{x^2-y^2} \sim \Delta_0 (\cos k_x - \cos k_y)$ and $d_{xy} \sim
\Delta_0 \sin k_x \sin k_y$, and three possible $s$-wave types
\cite{seo2008}: $s_{x^2+y^2} \sim \Delta_0 (\cos k_x + \cos k_y)$,
$s_{x^2y^2} \sim \Delta_0 \cos k_x \cos k_y$, as well as the
constant gap ($s_0$) which is not allowed in  the $t-J_1-J_2$ model
but can obviously appear in other interacting models. The $C_4$
symmetry of the underlying lattice maps $k_x \leftrightarrow k_y$
and $d_{xz} \leftrightarrow d_{yz}$. Hence for all the pairing
symmetries described above we have $\Delta_1(k_x,k_y) =
\Delta_2(k_y,k_x)$, except for $d_{x^2 - y^2}$ where $\Delta_1
(k_x,k_y) =-\Delta_2 (k_x,k_y)$. \cite{seo2008} The $d_{x^2-y^2}$,
$d_{xy}$ and $s_{x^2+y^2}$ pairing symmetries are nodal while the
other pairing symmetries are nodeless.  We now proceed to analyzing
the experimental consequences of these pairing symmetries.

\paragraph*{Spectral Function, Density of States and Tunneling Differential Conductance --}
The single-particle density of states (DOS) can be written as:
\begin{align}
& \mathcal{N}(\omega) \equiv \sum_{\vect{k}} \mathcal{A}(\vect{k},\omega) \nonumber \\
& = -\frac{1}{\pi} \sum_{\vect{k}}
\Im[\mathcal{G}_{11}(\vect{k},\omega+i\delta) +
\mathcal{G}_{33}(\vect{k},\omega+i\delta)]
\end{align}
where $\mathcal{A}(\vect{k},\omega)$ is the spectral function and
$\mathcal{G}_{11}(\vect{k},\omega+i\delta)$  and
$\mathcal{G}_{33}(\vect{k},\omega+i\delta)$ are the electron
components of the superconducting Green's function. Generally, we
find:
\begin{widetext}
\begin{align}\notag
 \mathcal{A}(\vect{k},\omega) = & \frac{\epsilon_{xy}^2 (2\omega -
 \xi_x - \xi_y) - (\omega + \xi_y) (\omega^2 - \xi_x^2 - \Delta_1^2)
 - (\omega + \xi_x) (\omega^2 - \xi_y^2 - \Delta_2^2) }
 {E_1^2 - E_3^2} \\
 & \times \left[\frac{1}{2E_3}(\delta(E_3 - \omega) - \delta(E_3 + \omega))
 - \frac{1}{2E_1} (\delta(E_1 - \omega) - \delta(E_1 + \omega)) \right]
\end{align}
where $E_1$ and $E_3$ are the positive eigenvalues of the matrix
$B(\vect{k})$ in \eqref{eq:Bmat} (see Ref.~\onlinecite{seo2008}).
For the case where $\Delta_1 = \Delta_2 = \Delta$ (valid except
for the $d_{x^2-y^2}$ pairing symmetry), we have the simplified
form:
\begin{align}
\mathcal{A}(\vect{k},\omega) = & \frac{\omega +
E_-(\vect{k})}{2E_-^\Delta(\vect{k})} [\delta(E_-^\Delta(\vect{k}) -
\omega) -
\delta(E_-^\Delta(\vect{k}) + \omega)] 
+ \frac{\omega + E_+(\vect{k})}{2 E_+^\Delta(\vect{k})}
[\delta(E_+^\Delta(\vect{k}) - \omega) - \delta(E_+^\Delta(\vect{k})
+ \omega)]
\end{align}
\end{widetext}
with $E_\pm^\Delta(\vect{k}) = \sqrt{E_\pm^2(\vect{k}) +
\Delta^2(\vect{k})}$. This resembles two independent single-band
superconductors with the energy dispersions $E_{\pm}$.

The spectral function at the Fermi energy
$\mathcal{A}(\vect{k},\omega = 0)$ contains information about the
nodal structure for each pairing symmetry, as shown in
Fig.~\ref{fig:Sfuncs}. The $s_{x^2+y^2}$ pairing symmetry exhibits
nodes on the Fermi surface for all dopings when $k_x = (\pm\pi -
k_y), (\pm\pi + k_y)$ and thus only the hole Fermi pockets are fully
gapped. The $d_{xy}$ pairing symmetry also has nodes for all doping,
but in this case they occur when $k_{x,y} = 0,\pm\pi$ and so all of
the Fermi surfaces are gapless. The $d_{x^2-y^2}$ pairing symmetry
(not shown) exhibits nodes on  the Fermi surface of the hole pockets
for any doping and it has a similar effect on the electron pockets
as the $s_{x^2y^2}$ pairing symmetry which is the dominant pairing
symmetry that we found in Ref.~\onlinecite{seo2008}. The
$s_{x^2y^2}$ pairing only has nodes on the Fermi surface above a
critical doping $\mu \simeq 2$ since the zeros of the gap lie at
$k_{x,y} = \pm \pi/2$. For $\mu < 2$, the electron Fermi surfaces
are fully gapped, like the hole Fermi surfaces. In principle,
information about the form of the $s_{x^2y^2}$ gap can be obtained
through  ARPES. In the \emph{folded} Brillouin Zone, there are two
hole pockets at the $\Gamma$ point. A $\cos(k_x)\cdot \cos(k_y)$
order parameter predicts a \emph{larger} gap for the smaller hole
Fermi surface and \emph{smaller} gap for the larger hole Fermi
surface.

\begin{figure}
\centering
\includegraphics[width = 0.48\textwidth]{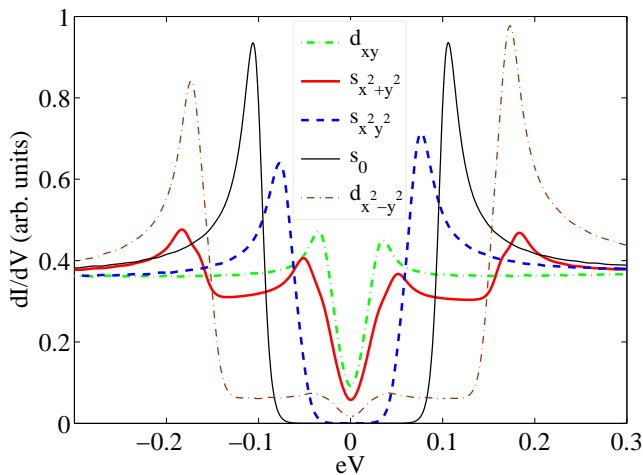}
\caption{\label{fig:DOS} Tunneling differential conductance $dI/dV
\propto -\int \mathcal{N}(\omega) n_F^\prime(\omega - eV)$ as a
function of bias voltage $eV$ measured with respect to the Fermi
energy, where the temperature $k_B T = 0.005 $, the chemical
potential $\mu=1.6$, and the gap size $\Delta_0 = 0.1$ for all the
different pairing symmetries.}
\end{figure}

\begin{figure}
\centering
\includegraphics[width = 0.46\textwidth]{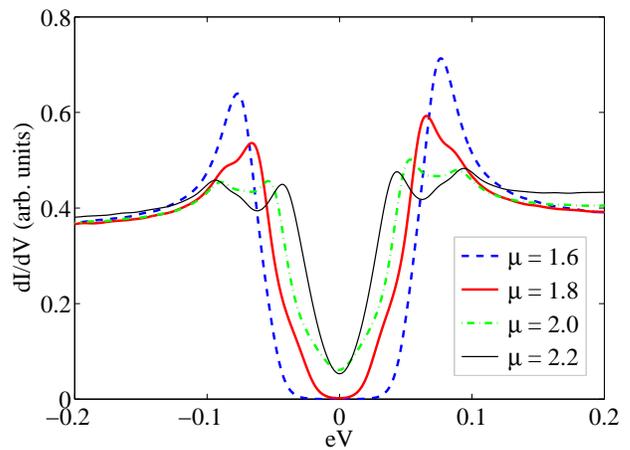}
\caption{\label{fig:DOS_swave} Tunneling differential conductance
$dI/dV$ as a function of bias voltage $eV$ for the pairing symmetry
$s_{x^2y^2}$ at different dopings. Like Fig.~\ref{fig:DOS}, $k_BT =
0.005$ and $\Delta_0 = 0.1$.}
\end{figure}

Tunneling measurements access the local DOS to a first
approximation. Specifically, if we assume that both the tunneling
matrix element and the probe DOS are momentum independent, then the
tunneling differential conductance is~\cite{fischer2007}:
\begin{align}
 \frac{dI}{dV} & \propto - \int^\infty_{-\infty}
 \mathcal{N}(\omega) n_F^\prime(\omega - eV)
\end{align}
where $eV$ is the bias voltage of the tunneling probe and
$n^\prime_F(E) \equiv \partial n_F(E) /\partial E$ is the derivative
of the Fermi function. In the limit of zero temperature, we
obviously recover the DOS. From Fig.~\ref{fig:DOS}, we see that a
fully-gapped Fermi surface yields a corresponding gap in $dI/dV$ at
low energies, while for a gapless Fermi surface, the DOS grows
quasi-linearly with energy at $\omega = 0$, a text-book result. In
particular, $s_{x^2+y^2}$ ($d_{x^2-y^2}$) pairing produces a
four-peak structure in the differential conductance because in this
case each band sees a different order parameter: the hole (electron)
Fermi surfaces are fully gapped while the electron (hole) Fermi
surfaces are gapless.

Focussing on $s_{x^2y^2}$ pairing\cite{seo2008}
(Fig.~\ref{fig:DOS_swave}), we find that the differential
conductance smoothly evolves from fully gapped to gapless behavior
with increasing doping, as expected. Moreover, when the doping is
large, we obtain a four-peak structure similar to $s_{x^2+y^2}$
pairing, because we also have a fully-gapped hole Fermi surface and
a partially-gapped electron Fermi surface. While it is likely that
the material cannot be doped high enough so that the  $s_{x^2 y^2}
=\cos(k_x) \cdot \cos(k_y)$ superconductor becomes gapless (the
material will most likely exit the superconducting state at such
high dopings), we believe that the predictions above, in particular
the evolution of the differential conductance with doping, could be
used in careful experiments to falsify this order parameter.

\paragraph*{Penetration depth --}
Measurements of the penetration depth in the Fe-based
superconductors were the first to suggest that the Fermi surfaces
are fully gapped~\cite{malone2008,martin2008,hashimoto2008}. The
experiments show an exponentially temperature decay of $\delta
\lambda(T) = \lambda(T) - \lambda(0)$. Among the different order
parameters studied here, such a scenario is only consistent with
$s_{x^2y^2}$ symmetry at low doping or a constant $s$-wave gap. We
now obtain the penetration depth for the bare two band
superconductor with generic $\Delta_{1,2}$ gaps.

To obtain the penetration depth, we perform a text-book exercise. We
write the FeAs model in real space and introduce a gauge field via
the Peierls substitution $c_{i,\alpha}^\dagger c_{j,\beta}
\rightarrow c_{i,\alpha}^\dagger \exp(i \int_{i}^j \vec{A} \cdot d
\vec{l}) c_{j,\beta}$ where $\alpha, \beta$ are the two orbital
indices. We pick a Landau gauge $\vec{A} = A \hat{x}$ and expand to
second order in $A$, thus obtaining $H(A)$. The second order term in
$A$ is the diamagnetic current  while the first order term gives the
paramagnetic current, whose response must be calculated in linear
response. We have:
\begin{equation}
H(A) \approx H(0) -\sum_i (j_x^p(i) A_x(i)  + \frac{1}{2} j_x^d(i) A_x(i)^2 )
\end{equation}
hence
\begin{equation}
j_x(i) = - \frac{\delta H(A)}{\delta A_x(i)} = j_x^p(i) + j_x^d(i) A_x(i)
\end{equation}
Using translational invariance, the expectation value of the
diamagnetic current in the ground state is:
\begin{multline}
\braket{j_x^d(i)} = \frac{1}{N_s} \sum_i \braket{j_x^d(i)} \\
 = - \frac{1}{V}
\sum_{\vect{k}} \frac{\partial^2 \epsilon_x}{\partial k_x^2}
\braket{c_{\vect{k},1}^\dagger c_{\vect{k},1}} +\frac{\partial^2
\epsilon_y}{\partial
k_x^2} \braket{c_{\vect{k},2}^\dagger c_{\vect{k},2}} \\
 + \frac{\partial^2
\epsilon_{xy}}{\partial k_x^2} \braket{c^\dagger_{\vect{k},1}
c_{\vect{k},2} + c^\dagger_{\vect{k},2} c_{\vect{k},1}}
\end{multline}
where the expectation values of the above operators are computed in
the appropriate ground state. The paramagnetic current is obtained
through a correlation function in linear response:
$j_x^p(\vect{q},\omega)=Q_{xx}(\vect{q}, \omega) A_x(\vect{q},
\omega)$:
\begin{equation}
Q_{xx}^d(\vect{q}, i \nu_n)  = \frac{1}{N} \int_0^\beta d\tau e^{i
\nu_n \tau} \braket{j_x^p(\vect{q},\tau) j_x^p(-\vect{q},0)}
\end{equation}
This is the vacuum polarization. For the FeAs metal (\emph{not} the
superconductor), this is explicitly given by:
\begin{align}
& Q_{xx}(\vect{q}, i \nu_n) = - \frac{1}{V\beta}\sum_{\vect{k},m}
\times  \nonumber \\
& Tr\left(J_x(\vect{k}) G(i\omega_m+ i \nu_n,
\vect{k}+\frac{\vect{q}}{2}) J_x(\vect{k}) G(i\omega_m,
\vect{k}-\frac{\vect{q}}{2})\right)
\end{align}
where $\omega_m = (2 m +1)\pi T$ is a fermionic Matsubara frequency
while $\nu_n= 2n \pi T$ is a bosonic one. $J_x$ is the current
operator, which is expressed as $\partial{H} /\partial{k_x}$ in the
metal. For the response to a magnetic field, the limit that has to
be taken is, upon analytic continuation, $i\nu_n \rightarrow
\omega+i \delta$, $\omega =0$, $ \vect{q}\rightarrow 0$. The
opposite limit $\omega \rightarrow 0$, $ \vect{q} = 0$ gives the
response to an electric field, and hence the electrical
conductivity. After tedious but straightforward algebra, we obtain
for the FeAs metal:
\begin{widetext}
\begin{equation}\label{eq:Qxx}
Q_{xx}(\vect{q} \rightarrow 0, \omega=0 ) =
 - \frac{2}{V} \sum_{\vect{k}} \left( \frac{\partial E_+}{\partial
k_x} \right)^2 \frac{\partial n(E_+)}{\partial E_+} + \left(
\frac{\partial E_-}{\partial k_x} \right)^2 \frac{\partial
n(E_-)}{\partial E_-}   + \frac{8(n(E_+) - n(E_-))}{(E_+ -
E_-)^3} \left(\epsilon_ {xy} \frac{ \partial \epsilon_-}{\partial
k_x} - \epsilon_- \frac{\partial \epsilon_{xy}}{\partial
k_x}\right)^2
\end{equation}
\end{widetext}
The overall factor of $2$ reflects the spin multiplicity. Besides
the usual paramagnetic expression (first two terms in
Eq.~\eqref{eq:Qxx}), the cross-orbital exchange introduces an extra
second term. We have checked that this paramagnetic term completely
cancels the diamagnetic ground state expectation value, as is
required for a metal. We performed the same calculation in the
superconductor.  The charge matrix operator in our superconductor
is:
\begin{equation}
J_0 = \left(
        \begin{array}{cccc}
          1 &0 & 0 &0 \\
          0 & -1 & 0 & 0 \\
          0 & 0 & 1 & 0 \\
          0 & 0 & 0 & -1 \\
        \end{array}
      \right)
\end{equation}
The current operator uses \emph{only} the kinetic part of the
kinetic Hamiltonian and is obtained from the continuity equation,
giving:
 \begin{equation}
J_x = \frac{1}{2} \left\{\frac{ \partial H(\Delta_1 =0, \Delta_2=0)}{\partial k_x}, J_0\right\}
\end{equation}
where $\{,\}$ is the anticommutator. The penetration depth $\delta
\lambda(T)= \lambda(T)- \lambda(0)$ is proportional to the
current-current correlation function which uses the Green's function
of the superconductor, not written here due to space restrictions.
For the case where $\Delta_1 = \Delta_2$, we can write the
current-current correlation function as:
\begin{widetext}
\begin{multline}
Q_{xx}(q\rightarrow 0,\omega = 0) =  -\sum_{\vect{k}} 2 \left[
\left(\frac{\partial E_+}{\partial k_x}\right)^2
n^\prime_F(E_+^\Delta) + \left(\frac{\partial E_-}{\partial k_x}
\right)^2 n^\prime_F(E_-^\Delta) \right]  \\
 + \frac{1}{\xi_+ (\xi_-^2 +
\epsilon_{xy}^2)^{3/2}} \left(\epsilon_{xy} \frac{\partial
\xi_-}{\partial k_x} - \xi_- \frac{\partial \epsilon_{xy}}{\partial
k_x} \right)^2 \times \left[(2n_F(E_+^\Delta) - 1)\frac{\xi_+ E_+ +
\Delta^2}{E_+^\Delta} - (2n_F(E_-^\Delta) - 1)\frac{\xi_+ E_- +
\Delta^2}{E_-^\Delta} \right]
\end{multline}
\end{widetext}
We see that the cross-orbital exchange introduces an extra term,
similar to the case of the FeAs metal, but the largest contribution
to the temperature dependence arises from the first term.  We have
obtained the expression of the current-current correlation function
for general $\Delta_1\ne \Delta_2$, but we do not include it for
space reasons.

\begin{figure}
\centering
\includegraphics[width = 0.48\textwidth]{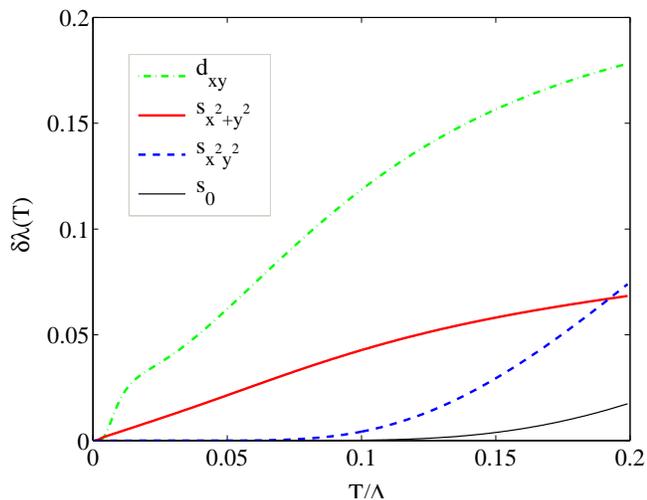}
\caption{Penetration depth $\delta\lambda (T) = \lambda(T) -
\lambda(0) \propto Q_{xx}(\vect{q} \rightarrow 0, \omega=0 )$ close
to zero temperature for different pairing symmetries at doping $\mu
= 1.6$ and gap size $\Delta_0 = 0.1$. The $d_{xy}$ curve has been
reduced by a factor of two for clarity. The $d_{x^2-y^2}$ pairing
symmetry (not shown) will have a similar low temperature behavior to
the $d_{xy}$ and $s_{x^2+y^2}$ curves. \label{fig:pendep}}
\end{figure}

\begin{figure}
\centering
\includegraphics[width = 0.48\textwidth]{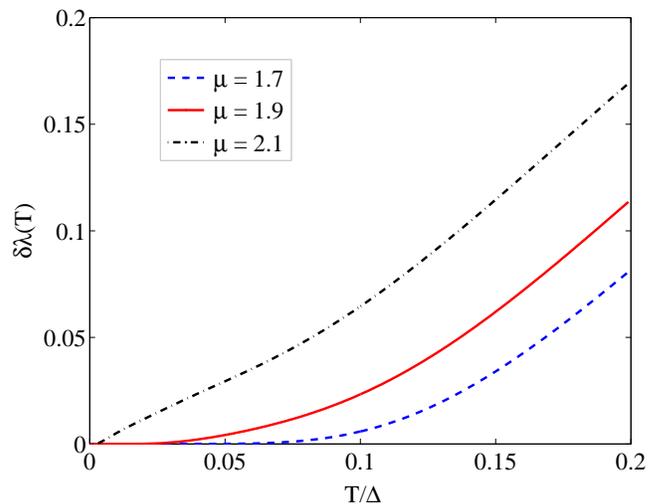}
\caption{Penetration depth $\delta\lambda (T) = \lambda(T) -
\lambda(0)$ for the pairing symmetry $s_{x^2 y^2}$ at different
dopings, where $\Delta_0 = 0.1$. \label{fig:pendepS}}
\end{figure}

We now plot the low temperature dependence of the penetration depth
$\delta \lambda(T) =\lambda(T) - \lambda(0)$ for different
superconducting gaps (see Fig.~\ref{fig:pendep}). As expected, the
nodal order parameters exhibit a linear $T$ dependence (in the
absence of impurities) while the nodeless order parameters exhibit
an exponentially decaying penetration depth. However, as shown in
Fig.~\ref{fig:pendepS}, one qualitative feature is that the
$\cos(k_x) \cdot \cos(k_y)$ order parameter exhibits, upon doping, a
weakened exponential decay, a signature that the gap on the electron
(hole) surface \emph{decreases} upon electron (hole) doping. This is
a direct consequence of the existence of a line of zeroes in between
the electron and hole pockets. Above some critical doping, the
exponential decay of $\delta \lambda(T)$ in the $\cos(k_x) \cdot
\cos(k_y)$ superconductor becomes linear (Fig.~\ref{fig:pendepS}), a
sign that the superconductor has become gapless.

\paragraph*{NMR Spin Relaxation Rate and the Coherence Peak --}
Existing experimental results for the NMR spin relaxation time $T_1$
at first sight suggest a $d$-wave symmetry for the order parameter,
because there is no coherence peak in $1/T_1$ at $T_c$ and $1/T_1$
scales like $T^3$ just below
$T_c$~\cite{matano2008,grafe2008,nakai2008,mukuda2008}. These
results pose a big challenge for the $s$-wave pairing symmetry or
any other nodeless order parameter. In the case of a $\cos(k_x)
\cdot \cos(k_y)$ order parameter, although we find that the
coherence peak due to inter-band contributions is non-existent, the
intra-band contributions still give a coherence peak, although
smaller and flatter than in a pure $s$-wave scenario. Neglecting the
intra-band contributions (which could be justified if the
broadenings of the inter- and intra-band contributions are
different) can then explain the observed lack of the coherence peak,
but in general a small coherence peak should be seen in cleaner
samples.

The NMR measurements have been performed on different atoms in the
pnictides, including $^{19}$F and $^{75}$As. Experimentally, there
is no major difference between the $1/T_1$ results on these two
atoms. This also poses a challenge to the NMR theories because the
structure factors for F and As are different: while the structure
factor for F is roughly isotropic in the transferred momentum
$\vect{q}$, the As structure factor is roughly $A(\vect{q}) =
\cos(q_x/2)\cos(q_y/2)$ due to the placement of the As atoms in the
center of the Fe unit cell (although the As are out of plane, we
believe the $\cos(q_x/2)\cos(q_y/2)$ faithfully represents the
structure factor). Hence, for small Fermi electron and hole pockets,
the As NMR measurements should not be sensitive to the inter-band
contributions, whose transfer wavevector $(\pi,0)$
is suppressed by the structure factor.

The NMR spin relaxation rate at temperature $T$ is defined as:
\begin{equation}
R = \frac{1}{T_1 T} =-\frac{1}{2 \pi} \lim_{\omega_0 \rightarrow 0 }
\frac{\Im[K_{+-}(\omega_0)]}{\omega_0}
\end{equation}
where
\begin{equation}
K_{+-}(\omega_0) = \sum_q A(\vect{q})\xi^{+-}(\vect{q}, \omega_0)
\end{equation}
$ \xi^{+-}(\vect{q}, \omega_0)$ is the spin susceptibility in the
superconducting state and $A(\vect{q})$ is the structure factor.
Since we are dealing with singlet superconductivity we have:
\begin{equation}
\xi^{+-} = \frac{1}{2}(\xi^{xx}+\xi^{zz}) = \xi^{zz}
\end{equation}
where $\xi^{zz}$ is now much simpler due to the fact that the $S^z$
spin matrix in a superconductor is the identity matrix:
\begin{multline}
K_{+-}(\omega_0)  = \frac{1}{V^2 \beta} \sum_{\omega_n,\vect{k_1},
\vect{k_2}} A(\vect{k_2}-\vect{k_1})  \\
 \times Tr[G({\vect{k_1},i(\omega_n+ \omega_0)})G({\vect{k_2},i\omega_n})]
\end{multline}
After Matsubara sums, analytic continuation, and taking the
imaginary part, for the pure gap case $\Delta_1 = \Delta_2 =
\Delta$, we obtain the following formula for $1/(T_1T)$,
\begin{widetext}
\begin{align}\notag
\frac{1}{T_1 T} = \sum_{\vect{k_1}, \vect{k_2}} &
A(\vect{k_2}-\vect{k_1}) \left[ \left( 1 + \frac{\Delta(\vect{k_1})
\Delta(\vect{k_2}) + E_+(\vect{k_1})
E_+(\vect{k_2})}{E_+^{\Delta}(\vect{k_1})^2} \right) \frac{\partial
n}{\partial
E_+^{\Delta}(\vect{k_1})} \delta(E_+^{\Delta}(\vect{k_2}) - E_+^{\Delta}(\vect{k_1})) \right. \\
\notag & + \left( 1 + \frac{\Delta(\vect{k_1}) \Delta(\vect{k_2}) +
E_-(\vect{k_1}) E_-(\vect{k_2})}{E_-^{\Delta}(\vect{k_1})^2} \right)
\frac{\partial n}{\partial E_-^{\Delta}(\vect{k_1})}
\delta(E_-^{\Delta}(\vect{k_2}) - E_-^{\Delta}(\vect{k_1})) \\
\label{eq:T1}
 & \left. + 2 \left( 1 + \frac{\Delta(\vect{k_1})
\Delta(\vect{k_2}) + E_+(\vect{k_1})
E_-(\vect{k_2})}{E_+^{\Delta}(\vect{k_1})^2} \right) \frac{\partial
n}{\partial E_+^{\Delta}(\vect{k_1})}
\delta(E_-^{\Delta}(\vect{k_2}) - E_+^{\Delta}(\vect{k_1}))\right]
\end{align}
\end{widetext}
The first two terms in Eq.~\eqref{eq:T1} represent the intra-band
contribution and the third term represents the inter-band
contribution, which is a contribution between the electron and hole
pockets.
Following Bulut and Scalapino,~\cite{bulut1992} we
phenomenologically take disorder into consideration by broadening
the Kronecker delta functions, e.g.\ $\pi
\delta(E_-^{\Delta}(\vect{k_2}) - E_+^{\Delta}(\vect{k_1}) ) = -
\Gamma/((E_-^{\Delta}(\vect{k_2}) -
E_+^{\Delta}(\vect{k_1})^2+\Gamma^2)$. This simple inclusion of
disorder works well towards explaining the experimental data in the
cuprate case, and merely serves as a cutoff for the singularities in
the density of states. We perform the momentum integrals by Monte
Carlo evaluation: this is necessary due to the fact that we keep the
strong-coupling superconductivity and do not make the usual
approximation which transforms the 4 momentum integrals and the
delta function into an easy one dimensional integral over energies
close to the Fermi surface.

\begin{figure}
\centering
\includegraphics[width = 0.48\textwidth]{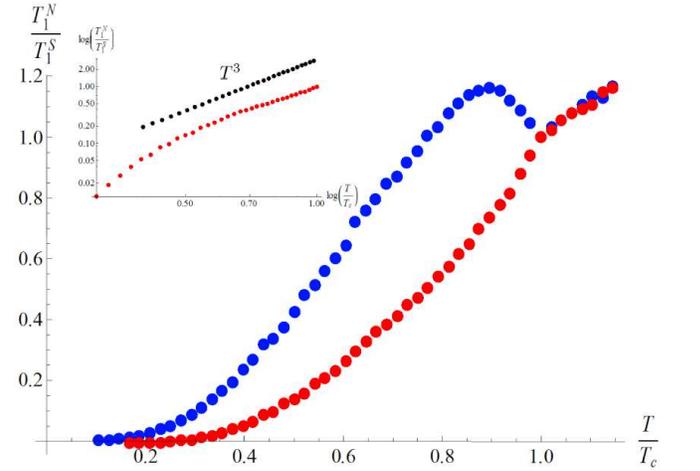}
\caption{Monte-Carlo calculation of the (normalized) inter-band
contributions to  the NMR coherence peak for $\Delta_0 \cos(k_x)
\cdot \cos(k_y)$ (red) and a fixed-sign version of it, $\Delta_0|
\cos(k_x) \cdot \cos(k_y)|$ (blue). We choose a large $\Delta_0=
|t_1|/5$ and $\Delta_0/T_c=2$. The broadening factor is $\Gamma =
T_c/5$, and $\mu=1.8$ corresponding to $18\%$ electron doping.
Inset: Temperature dependence of the NMR spin relaxation time for
$\Delta_0 \cos(k_x) \cos(k_y)$ (red). The structure factor here is
taken to be $A(q) =1$. \label{fig:nmr_intercos}}
\end{figure}

\begin{figure}
\centering
\includegraphics[width = 0.48\textwidth]{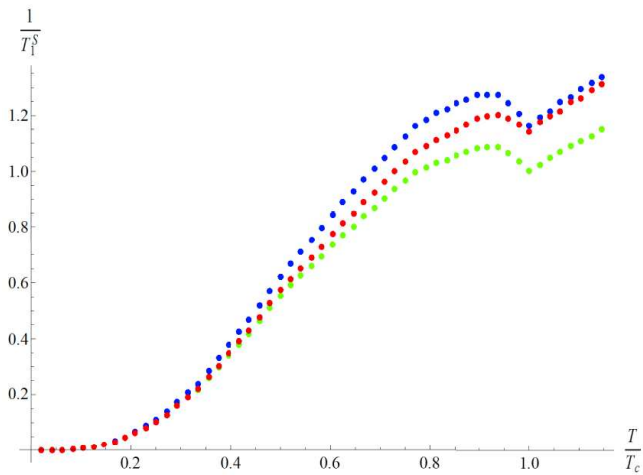}
\caption{Monte-Carlo calculation of the (normalized) intra-band
contributions to  the NMR coherence peak for the $\Delta_0 \cos(k_x)
\cdot \cos(k_y)$ (green) gap (the intra-band contribution is equal
for the two gaps $\Delta_0 \cos(k_x) \cdot \cos(k_y)$ and $\Delta_0|
\cos(k_x) \cdot \cos(k_y)|$). The total, intra plus inter band
contributions for $\Delta_0 \cos(k_x) \cdot \cos(k_y)$ (red) and
$\Delta_0 |\cos(k_x) \cdot \cos(k_y)|$ (blue) are also plotted. We
can see that the intra-band contribution is hence much larger than
the inter-band contribution for both these order parameters, and
hence the $\Delta_0 \cos(k_x) \cdot \cos(k_y)$ gap should exhibit a
small coherence peak. We choose a large $\Delta_0= |t_1|/5$ and
$\Delta_0/T_c=2$. The broadening factor is $\Gamma = T_c/5$, and
$\mu=1.8$ corresponding to $18\%$ electron doping.  The structure
factor here is taken to be $A(q) =1$. \label{fig:nmr_costot}}
\end{figure}

The inter-band and intra-band contributions have different behaviors
as a function of temperature.  Owing to the fact that for
$\vect{k_1}$ on the hole Fermi surface and $\vect{k_2}$ on the
electron Fermi surface, $\Delta(\vect{k_1})>0$ while
$\Delta(\vect{k_2})<0$, we expect the inter-band contribution to
lack a coherence peak around the superconducting transition
temperature, which is indeed what we find below.

We first consider a uniform structure factor, i.e.\ $A(\vect{q})=1$.
In Fig.~\ref{fig:nmr_intercos}, we contrast the inter-band
contribution for the $s_{x^2y^2}$ pairing symmetry with that of its
absolute value, i.e.\ $|\cos(k_x) \cdot \cos(k_y)|$, which does not
exhibit a sign change between the hole and electron pockets.
Clearly, the former case does not possess a coherence peak, while
the latter does, as expected.
In Fig.~\ref{fig:nmr_costot}, we plot the intra-band contribution
and the total $1/T_1$ for both cases. We see that, compared to the
absolute value case, the coherence peak in $1/T_1$ is suppressed in
the $\cos(k_x) \cdot \cos(k_y)$ case.

\begin{figure}
\centering
\includegraphics[width = 0.48\textwidth]{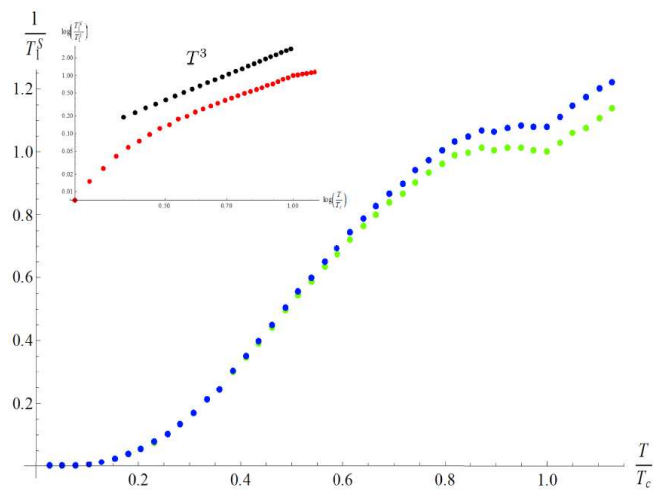}
\caption{Monte-Carlo calculation of the (normalized) intra-band
contributions to the NMR coherence  peak for the $\Delta_0 \cos(k_x)
\cdot \cos(k_y)$ (green) gap, with the structure factor
$A(\vect{q})=\cos(q_x/2)\cos(q_y/2)$ for the As atoms. The total,
intra plus inter band contributions for $\Delta_0 \cos(k_x) \cdot
\cos(k_y)$ (blue) are also plotted. The coherence peak is diminished
from the case when $A(\vect{q})=1$, plotted previously. Inset:
Temperature dependence of the inter band contribution. We choose a
large $\Delta_0= |t_1|/5$ and $\Delta_0/T_c=2$. The broadening
factor is $\Gamma = T_c/5$, and $\mu=1.8$ corresponding to $18\%$
electron doping. \label{fig:nmr_cos_strucA}}
\end{figure}

Using the structure factor $A(\vect{q})$ for As atoms
(Fig.~\ref{fig:nmr_cos_strucA}) we find that the inter-band
component of the total NMR spin relaxation rate \emph{decreases}.
While for $A(\vect{q})=1$ the inter-band contribution represented
about $1/6$ of the overall spin relaxation rate, for
$A(\vect{q})=\cos(q_x/2)\cos(q_y/2)$ that ratio decreases to about
$1/12$. We hence find that the intra-band contribution is dominant
in the case of the As structure factor. However, we also find that
the structure factor \emph{reduces the intra-band coherence peak},
to give an overall result plotted in Fig.~\ref{fig:nmr_cos_strucA}.

\begin{figure}
\centering
\includegraphics[width = 0.48\textwidth]{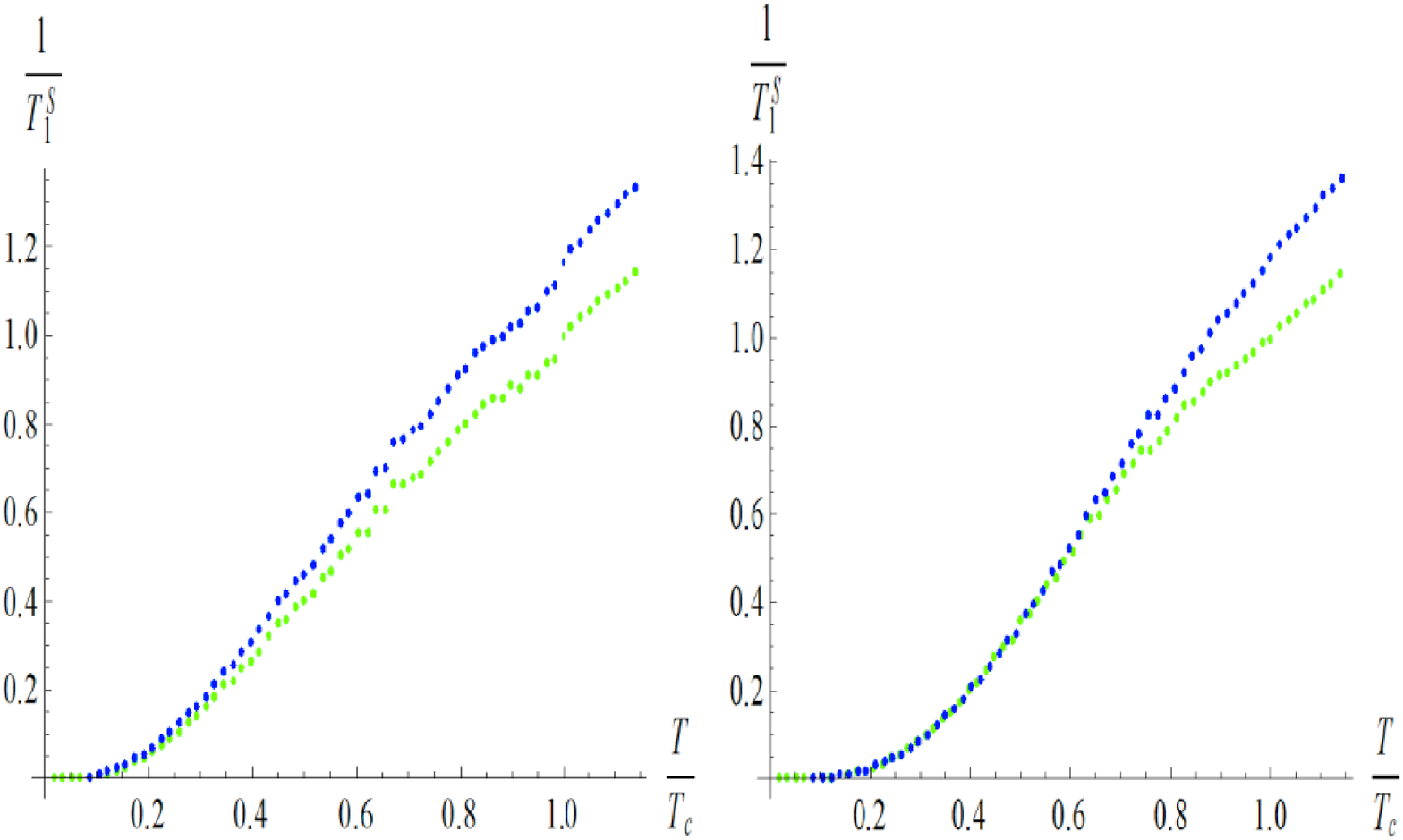}
\caption{Monte-Carlo calculation of the (normalized) intra-band
(green) and total (blue)  contributions to the NMR coherence peak
for the $\Delta_0 \sin(k_x)  \sin(k_y)$ (left) and $\Delta_0
(\cos(k_x) + \cos(k_y))$  gap. These are nodal superconductors and
lack a coherence peak. We choose a large $\Delta_0 = |t_1|/5$ and
$\Delta_0/T_c=2$. The broadening factor is $\Gamma = T_c/5$, and
$\mu=1.8$ corresponding to $18\%$  electron doping. The structure
factor here is taken to be $A(q) =1$. \label{fig:nmr_sincos}}
\end{figure}

Finally, we find that the NMR relaxation rates for the nodal
superconductors $d_{xy}$ and $s_{x^2 + y^2}$, depicted in
Fig.~\ref{fig:nmr_sincos}, lack a coherence peak as expected.

We predict that future experiments will see a small coherence peak
resulting from the intra-band contribution. Our results show that,
barring different scattering rates for inter- and intra-band
scattering, the overall intra-band contribution to the NMR
relaxation rate is roughly a factor of $5$ times larger than the
inter-band contribution. This can also be argued on general grounds,
provided that the hypothesis of weak-coupling theories and LDA
(i.e.\ there is a quasi-nesting of the electron and hole Fermi
surfaces in the parent material) is correct. Upon doping with either
electrons or holes, either the electron or hole Fermi surfaces will
become considerably larger than the other one. This means that the
inter-band contribution to the NMR spin relaxation rate diminishes:
it of course vanishes if one could, theoretically, deplete one of
the Fermi pockets. Meanwhile, the intra-band contribution should, on
general grounds, remain roughly constant upon doping because the
overall size of the sum of the Fermi surfaces is relatively
constant. All these general arguments are supported by our explicit
calculation.

A few other remarks about the NMR spin relaxation rates are in
order: (i) The observed $T^3$ temperature dependence of $1/T_1$
cannot be viewed as evidence against $s$-wave pairing symmetries. In
fact, the temperature dependence just below $T_c$ is very sensitive
to the ratio $\Delta/k_B T_c$. We find that the $T^3$ behavior can
be obtained by choosing $\Delta/k_B T_c \sim 2 $ for our large gap
value, and the power of the temperature dependence can increase even
further by increasing this ratio.  (ii) Although we predict that
there should be a coherence peak in the clean limit, impurities can
efficiently reduce the coherence peak in a two-band system. A weak
inter-band impurity scattering but strong intra-band scattering can
suppress the coherence peak.  This has been investigated in
MgB$_2$\cite{Mitrovic2006} where the coherence peak is also not
easily observed experimentally~\cite{Kotegawa2001}. Since the
superconductivity in Fe-based superconductors is created by doping,
it is reasonable to assume that disorder is stronger than that in
MgB$_2$. To observe the coherence peak, we require a very clean
sample. (iii) Our calculation is based on a two-band model. This
model can be over-simplified when one tries to use it to predict
quantitative experimental measurements. For example, the detailed
shape of  Fermi surfaces and its doping dependence may not be
quantitatively accurate. Therefore, the predictions in this paper
with regard to doping concentration should be viewed as qualitative.

\paragraph*{Conclusion --}
We have calculated the spectral function, the DOS, the tunneling
differential conductance, the penetration depth, and the NMR spin
relaxation rate for different superconducting order parameters in
the iron-pnictides. We have emphasized that the nodal structure of
the $s_{x^2y^2}$ order parameter will result in a qualitative change
in these experimental observables with increasing doping, as the
superconductor crosses over from gapped to gapless. Thus, one can in
principle probe the existence of this pairing symmetry in the
iron-pnictides by analyzing the behavior of the spectral function,
the DOS and the penetration depth as a function of doping.  For the
$1/T_1$ NMR spin relaxation rate, if only the inter-band
contribution is considered, our theoretical results are consistent
with the current experimental results, including the $T^{3}$
behavior and the absence of a coherence peak. However, by including
the intra-band contribution, a small coherence peak at the
transition temperature will be present in a clean sample although it
is smaller than that in a sign-unchanged $s$-wave.

\textit{Note --} During the completion of this work, we became aware
of two recent papers that also calculate the spin-lattice relaxation
rate for the $s_{x^2y^2}$ order parameter in the
iron-pnictides~\cite{parker2008,chubukov2008}, and another recent
paper that considers the experimental consequences of two different
pairing symmetries~\cite{bang2008}.

\paragraph*{Acknowledgements}
BAB wishes to thank P.W. Anderson, S. Sondhi, N.P. Ong, Z. Hasan, A.
Yazdani, for discussions and comments.  MMP and BAB are especially
grateful to David Huse for fruitful discussions. JPH thanks
Pengcheng Dai, S. Kivelson,  X. Tao,  E. W. Carlson, G. Q. Zheng,
I.I. Mazin, H. Yao,  for important discussion.  JPH was supported by
the National Science Foundation under grant No. PHY-0603759. MMP and
BAB are supported by PCTS fellowships.


\end{document}